\documentclass[10pt]{article}
\usepackage{hyperref}
 \usepackage{footnote}
\makesavenoteenv{minipage}
\renewcommand{\title}[1]{%
    \bigskip%
    \begin{center}%
    \Large\bf #1%
    \end{center}%
    \vskip .2in}

\renewcommand{\author}[1]{%
    {\begin{center}
    #1
    \end{center}}}
\newcommand{\address}[1]{\vspace{-1.7em}\vspace{0pt}
    {\begin{center}
    \it #1 
    \end{center}}}

\begin{document}
\title{Finite field-dependent BRST symmetry for  ABJM theory in ${\cal N}=1$ superspace}
\author{ Mir Faizal\footnote{ f2mir@uwaterloo.ca }}
\address{Department of Physics and Astronomy,   University of Waterloo,   Waterloo,\\
Ontario N2L 3G1, Canada }
\author{Sudhaker Upadhyay\footnote{sudhakerupadhyay@gmail.com} and  Bhabani Prasad Mandal\footnote{bhabani.mandal@gmail.com}}\address{ Department of Physics, Banaras Hindu University,  \\
Varanasi-221005, India}

\begin{abstract}
In this paper  we  
analyse the ABJM theory in ${\cal N}=1$ superspace. 
 Firstly we study the linear and non-linear  BRST transformations for the ABJM theory. 
Then we   derive the
finite field dependent version of these
BRST (FFBRST) transformations.  Further we show that
 such FFBRSTtransformations
relate the generating functional  
in  linear gauge to the generating functional in the non-linear gauge of ABJM theory. 
\end{abstract}

\section{Introduction}

According to the $AdS/CFT$ correspondence this superconformal field theory  is dual to
the eleven dimensional supergravity  on $AdS_4 \times S_7$.
Apart from a constant closed 7-form on $S^7$,
$AdS_4 \times S_7 \sim SO(2,3)\times SO(1,2)/ SO(8)\times SO(7) \subset OSp(8|4)/SO(1,3) \times SO(7)$. 
So, the dual superconformal field theory to 
  the eleven dimensional supergravity  on $AdS_4 \times S_7$
  has  $OSp(8|4)$ realized as $\mathcal{N} = 8$ supersymmetry. 
This theory also has  eight gauge valued scalar fields,  sixteen physical 
fermions and the gauge fields of this theory do not have any on-shell degrees of freedom. 
All these properties are satisfied by a theory called the BLG theory \cite{1, 2, 3, 4, 5}. 
The BLG theory is based on gauge symmetry generated by a Lie 3-algebra rather than a Lie algebra. 
So, far the only know example of a Lie 3-algebra is $SO(4) \sim SU(2) \times SU(2)$, and 
it corresponds to two M2-branes. It has not been possible to 
increase the rank of the gauge group. 

It has been possible to construct a superconformal gauge theory called the ABJM theory \cite{abjm, ab, ab1, 
ab2}.  
The ABJM theory only has 
 $\mathcal{N} = 6$ supersymmetry. However, it considers with the $BLG$ theory
for the only known example of the Lie 3-algebra and so its supersymmetry is expected to get enhanced to 
full $\mathcal{N} = 8$ supersymmetry \cite{abjm2}.
 The gauge sector is described by two Chern-Simons theories with 
levels $k$ and $-k$. The matter fields in the ABJM theory 
are in the bi-fundamental representation of the gauge group $U(N)_k \times U(N)_{-k}$
and the gauge fields are in the adjoint representation.
The ABJM theory has been studied in $\mathcal{N} =1$  and $\mathcal{N} =2$ superspace formalism \cite{6, 7, 
8}. 
The ABJM theory has also been studied in harmonic superspace \cite{ahs, mir4}. 
However, in this paper, we will analyse the ABJM theory in $\mathcal{N} =1$ superspace formalism. 
The 
  BRST  and the anti-BRST symmetries for the ABJM theory 
  have been studdied in both linear and non-linear gauges \cite{abm}. 

The infinitesimal  BRST 
transformations have been generalized 
to finite field dependent BRST (FFBRST) originally in \cite{jm} and further generalized to construct finite field dependent anti-BRST 
(FFanti-BRST) transformations in \cite{antifbrst} 
Similar generalizations have also been made recently in \cite{lav,lav1}.
This is done 
by first  making the infinitesimal global parameter occurring in the 
BRST or the anti-BRST transformations depend on fields occurring in the theory. Then this 
field dependent parameter  is integrated to obtain the FFBRST and anti-FFBRST transformations. 
Even though, these finite transformations  are a symmetry of the quantum action, they are not a symmetry of the 
 functional measure. They can thus be used to relate a theory 
in one gauge to the same theory in a different gauge \cite{lav}-\cite{ff1}. 
So, FFBRST transformations can be used to 
overcome  a problem that a 
theory suffers from in a particular gauge. This can be done by first calculating the required quantity in a 
gauge in which that problem does not exist, and then using the FFBRST transformation to 
transform it to the required gauge.
 Thus, 
in Yang-Mills theory, FFBRST transformations  have been used for obtaining the propagator in Coulomb gauge 
  from the generating function  in the Lorentz gauge \cite{ffb1}.
  The gauge-fixing and ghost terms corresponding to Landau and maximal Abelian gauge for the  
  Cho-Faddeev-Niemi  decomposed $SU(2)$ theory have also been generated using FFBRST transformation
  \cite{sud}. However, the linear and non-linear gauges of perturbative quantum gravity are connected at both classical and quantum level through FFBRST formulation \cite{sud1}.
  The quantum gauge freedom described by gaugeon formalism has also been studied 
  for quantum gravity \cite{sud2} as well as for Higgs model \cite{sud3} utilizing FFBRST technique.
The FFBRST transformations are also studied in the context of lattice gauge theory \cite{rbs1} and 
relativistic point particle model \cite{rbs}.

The  FFBRST transformation is used to relate the  Gribov-Zwanziger theory to Yang-Mills theory in Landau gauge
\cite{gz}. The problem of  formulating the Gribov-Zwanziger theory beyond the Landau gauge
is very delicate matter and substantial progress has been made recently towards the study
of this problem \cite{lav2,lav3}. 
Thus, FFBRST transformations may give us an idea about the non-perturbative 
effects in a theory. This is very important from the M-theory point of view. 
This is because we may be able to understand  the physics of multiple 
M5-branes by analysing non-perturbation effects in the ABJM theory \cite{m5}-\cite{m512}.  
The FFBRST transformations for the BLG theory    has already been  studied 
\cite{fs}.
However, this limits the analysis to two M2-branes. If we want to analyse similar effects 
for multiple M2-branes, we need to analyse a similar system for ABJM theory. 
It may be noted that in analysing the FFBRST symmetry for the ABJM theory, we will need to introduce 
two finite field dependent paremeters, which correspond to the gauge symmetries generated by $U(N)_k \times U(N)_{-k}$. 
As the matter fields transform in bi-fundamental representation of this gauge group, the matter sector 
mixes these two finite field dependent paremeters. Thus, we need to generalize the ordinary FFBRST symmetry, 
to apply it on the ABJM theory. This is what we aim to do in this paper.

The paper is organized as follows. In Sec. 2, we discuss the preliminaries about ABJM theory
in ${\cal N}=1$ superspace. The BRST symmetry for various gauges are presented in Sec. 3. The
FFBRST transformation for ABJM theory is developed in Sec. 4.
In Sec. 5, we relate two arbitrary gauges of ABJM theory using FFBRST transformation.
\section{ABJM Theory in ${\cal N}=1$ Superspace}
In this section we analyse ABJM theory on ${\cal N}=1$ superspace. For this purpose, we begin with the  Chern-Simons Lagrangian densities $\mathcal{L}_{CS}$, $\tilde{\mathcal{L}}_{CS}$ with 
gauge group's $U(N)_k$ and $U(N)_{-k}$ on ${\cal N}=1$ superspace defined by
\begin{eqnarray}
 \mathcal{L}_{CS}& =& \frac{k}{2\pi} \int d^2 \,  \theta \, \, 
 \mbox{Tr} \left[  \Gamma^a     \omega_a + \frac{i}{3} [\Gamma^a, \Gamma^b]_{ }
  D_b \Gamma_a  
 + \frac{1}{3} [\Gamma^a,\Gamma^b]_{ } [\Gamma_a, \Gamma_b]_{ }
\right]   , 
\nonumber \\
 \tilde{\mathcal{L}}_{CS} &=& -\frac{k}{2\pi} \int d^2 \,  \theta \, \, 
 \mbox{Tr} \left[  \tilde{\Gamma}^a     \tilde{\omega}_a
 + \frac{i}{3} [\tilde{\Gamma}^a, \tilde{\Gamma}^b]_{ }
  D_b \tilde{\Gamma}_a
+ \frac{1}{3} [\tilde{\Gamma}^a,\tilde{\Gamma}^b]_{ } 
[\tilde{\Gamma}_a, \tilde{\Gamma}_b]_{ }\right], 
\end{eqnarray}
where $\omega_a$ and $\tilde \omega_a$ have following expression:
\begin{eqnarray}
 \omega_a &=&  \frac{1}{2} D^b D_a \Gamma_b - i  [\Gamma^b, D_b \Gamma_a]_{  } 
- \frac{2}{3} [ \Gamma^b ,
[ \Gamma_b, \Gamma_a]_{ }]_{ },\nonumber \\
   \tilde\omega_a &=&  \frac{1}{2} D^b D_a \tilde\Gamma_b -i  [\tilde\Gamma^b, D_b \tilde\Gamma_a]_{  } - 
\frac{2}{3} [ \tilde\Gamma^b,
[ \tilde\Gamma_b,  \tilde\Gamma_a]_{  } ]_{  },
\end{eqnarray}
with the   super-derivative  $D_a$   defined by
$
 D_a = \partial_a + (\gamma^\mu \partial_\mu)^b_a \theta_b.
$

In the component form the super-gauge connections  $\Gamma_a$ and $\tilde \Gamma_a$ are described by 
\begin{eqnarray}
 \Gamma_a = \chi_a + B \theta_a + \frac{1}{2}(\gamma^\mu)_a A_\mu + i\theta^2 \left[\lambda_a -
 \frac{1}{2}(\gamma^\mu \partial_\mu \chi)_a\right], \nonumber \\
 \tilde\Gamma_a = \tilde\chi_a + \tilde B \theta_a + \frac{1}{2}(\gamma^\mu)_a \tilde A_\mu + i\theta^2 \left[\tilde \lambda_a -
 \frac{1}{2}(\gamma^\mu \partial_\mu \tilde\chi)_a\right]. 
\end{eqnarray}
The  Lagrangian density of the matter fields  is given by 
\begin{eqnarray}
 \mathcal{L}_{M} =\frac{1}{4} \int d^2 \,  \theta \, \,  
 \mbox{Tr} \left[ [\nabla^a_{(X)}           X^{I \dagger}           
\nabla_{a (X)}           X_I ] +
[\nabla^a_{(Y)}           Y^{I \dagger}           \nabla_{a (Y)}            Y_I ] + 
 \frac{16\pi}{k} \mathcal{V}_{    } \right]  ,
\end{eqnarray}
where 
\begin{eqnarray}
 \nabla_{(X)a}          X^{I } &=& 
D_a  X^{I } + i \Gamma_a          X^I 
-     i   X^I    \tilde\Gamma_a  , \nonumber \\ 
 \nabla_{(X)a}          X^{I \dagger} 
&=& D_a  X^{I  \dagger}      
+ i \tilde\Gamma_a        X^{I  \dagger}- 
     i X^{I  \dagger}   \Gamma_a, \nonumber \\ 
 \nabla_{(Y)a}          Y^{I } &=& D_a  Y^{I }  
+ i \tilde\Gamma_a          Y^I- i  Y^I   \Gamma_a , \nonumber \\ 
 \nabla_{(Y)a}          Y^{I \dagger} &=& D_a  Y^{I  \dagger} 
+ i \Gamma_a     
       Y^{I  \dagger} 
 - i   Y^{I  \dagger}     \tilde\Gamma_a.
\end{eqnarray}
Now, the  gauge invariant Lagrangian density for ABJM theory with the gauge group $U(N)_k \times U(N)_{-k} $ 
  on ${\cal N}=1$ superspace  is given by, 
\begin{equation}
{ \mathcal{L}_c} =  \mathcal{L}_{M} + \mathcal{L}_{CS} - \tilde{\mathcal{L}}_{CS}.
\end{equation} 
The  gauge transformations are given by
\begin{eqnarray}
\delta  \,\Gamma_{a} = \nabla_a   \xi, && \delta \, \tilde\Gamma_{a} =\tilde\nabla_a    
 \tilde \xi, \nonumber \\
\delta  \, X^{I } = i \xi X^{I }    -  iX^{I }  \tilde \xi, 
 &&  \delta  \, X^{I \dagger }
 = i   \tilde \xi  X^{I \dagger }  - i  X^{I \dagger } \xi, 
\nonumber \\
\delta  \, Y^{I } = i  \tilde \xi  Y^{I }  -iY^{I }  \xi,  &&  
\delta  \, Y^{I \dagger } = i \xi  Y^{I \dagger } -  i Y^{I \dagger }
  \tilde\xi,
\end{eqnarray}
with the local parameters $\xi$ and $\tilde\xi$. Here, the super-covariant derivatives  $\nabla_a$ and $\tilde\nabla_a$ are defined by
\begin{equation}
\nabla_a =D_a-i\Gamma_a, \ \ \ \tilde\nabla_a =D_a-i\tilde\Gamma_a.
\end{equation}
Not all the degrees of freedom of this      theory are physical as it 
 is invariant under gauge transformations.
\section{BRST Symmetry} 
In this section we will review the BRST symmetry for the  ABJM theory in the
 ${\cal N}=1$ superspace. 
 Being gauge invariant,  ABJM theory cannot be quantized  without 
 getting rid of these unphysical degrees of freedom. This is done by 
fixing the following gauge,
 \begin{eqnarray}
G_1 \equiv D^a \Gamma_a  =0,\ \  \tilde G_1 \equiv D^a \tilde{\Gamma}_a =0.
\end{eqnarray}
These gauge fixing conditions are incorporated at a quantum level by 
 adding  a gauge fixing term  $\mathcal{L}_{gf   }$ and a ghost term $\mathcal{L}_{gh   }$ 
 to the original classical Lagrangian. 
 Here the gauge fixing term is given by 
\begin{equation}
\mathcal{L}_{gf} = \int d^2 \,  \theta \, \,  \mbox{Tr}  \left[ib    (D^a \Gamma_a) + \frac{\alpha}{2}b  b  -
i \tilde{b}    (D^a \tilde{\Gamma}_a) - \frac{\alpha}{2}\tilde{b}     \tilde{ b} 
\right]  ,
\end{equation}
where $b $ and $\tilde b $ are the Nakanishi-Lautrup auxiliary fields. The Faddeev-Popov ghost term is given by  
\begin{equation}
\mathcal{L}_{gh} = \int d^2 \,  \theta \, \,   \mbox{Tr}
\left[ i\bar{c}    D^a \nabla_a    c  -i \tilde{\bar{c}}     D^a \tilde{\nabla}_a   \tilde{c}  \right]  .
\end{equation} 
The sum of the original Lagrangian density with the gauge fixing 
and ghost terms is invariant under the following BRST transformations  

\begin{eqnarray}
\delta_b \,\Gamma_{a} = \nabla_a    c \ \Lambda, && \delta_b\, \tilde\Gamma_{a} =\tilde\nabla_a    
 \tilde c \  \tilde \Lambda, \nonumber \\
\delta_b\,c  = - {[c ,c ]}_ { } \Lambda, && \delta_b \,\tilde{ {c}}  = -   [\tilde{ {c}} ,  \tilde c ]_{ } \tilde \Lambda, \nonumber \\
\delta_b \,\bar{c}  = b \ \Lambda, && \delta_b \,\tilde {\bar c}  = \tilde b \  \tilde \Lambda, \nonumber \\ 
\delta_b \,b  =0, &&\delta_b \, \tilde b = 0, \nonumber \\ 
\delta_b \, X^{I } = i  \Lambda c    X^{I }  -  iX^{I }  \tilde c  \tilde \Lambda, 
 &&  \delta_b \, X^{I \dagger }
 = i  \tilde \Lambda  \tilde c     X^{I \dagger } - i  X^{I \dagger }  c \ \Lambda, 
\nonumber \\
\delta_b \, Y^{I } = i \tilde \Lambda \tilde c    Y^{I }  -iY^{I }   c \ \Lambda,  &&  
\delta_b \, Y^{I \dagger } = i \Lambda c    Y^{I \dagger }\ -  i Y^{I \dagger }
  \tilde c \ \tilde \Lambda,
  \label{brstl}
\end{eqnarray}
where $\Lambda$ and $ \tilde \Lambda$ are the infinitesimal
anticommuting parameters of transformation. 

Now  we analyse ABJM theory in non-linear gauge and therefore we 
 define the Lagrangian density as follows 
  \begin{eqnarray} 
 {\cal L}_{NL}   &=& {\cal L}_c+ \int d^2\theta \ \mbox{Tr} \bigg[ \frac{\alpha}{2}  b ^2  + i b   D^a \Gamma_a  -iD^a \bar{c} \nabla_ac    - \frac{i}{2} D^a \Gamma_a [\bar{c} ,c ]  \nonumber \\  &+&    \frac{\alpha}{8} [\bar{c} ,c ]^2 - \frac{\alpha}{2}b  [\bar{c} ,c ]   +  iD^a \tilde{\bar{c}} \nabla_a\tilde{c}  - \frac{\alpha}{2} \tilde{b} ^2   - i\tilde{b}   D^a \tilde{\Gamma}_a  \nonumber \\ 
 &+&  \frac{i}{2} D^a \tilde{\Gamma}_a [\tilde{\bar{c}} ,\tilde{c} ]- \frac{\alpha}{8} [\tilde{\bar{c}} ,\tilde{c} ]^2 +\frac{\alpha}{2} \tilde{b} [\tilde{\bar{c}} , \tilde{c} ] \bigg]  .\label{lag}
\end{eqnarray}
We notice that the above Lagrangian density can be obtained by  shifting the  Nakanishi-Lautrup auxiliary fields as follows
 \begin{equation}
 b  \rightarrow b -\frac{1}{2} [\bar c , c ], \ \  \tilde b  \rightarrow \tilde b -\frac{1}{2} [\tilde{\bar c }, \tilde c ].
 \end{equation}
The BRST transformation, under which the effective action in non-linear gauge (\ref{lag})
is invariant, is given by  
\begin{eqnarray}
\delta_b \,\Gamma_{a} = \nabla_a    c \ \Lambda, && \delta_b\, \tilde\Gamma_{a} =\tilde\nabla_a    
 \tilde c \ \tilde\Lambda, \nonumber \\
\delta_b \,c  = -\frac{1}{2} {[c ,c ]}_ { }\ \Lambda, && \delta_b \,\tilde{ {c}}  = - \frac{1}{2}  [\tilde{ {c}}  ,  \tilde c ]_{ } \ \tilde\Lambda, \nonumber \\
\delta_b \,\bar{c}  = b \ \Lambda -\frac{1}{2}[\bar c , c ] \Lambda, && \delta_b \,\tilde {\bar c}  = \tilde b \ \tilde\Lambda -\frac{1}{2}[\tilde {\bar c} , \tilde c ]\ \tilde\Lambda, \nonumber \\ 
\delta_b \,b  =-\frac{1}{2}[c , b ]\Lambda- \frac{1}{8}[  [c ,c ] , \bar c ]\Lambda, &&\delta_b \, \tilde b = -\frac{1}{2}[\tilde c , \tilde b ] \tilde\Lambda- \frac{1}{8}[ [\tilde c , \tilde c ], \tilde{\bar c }]\tilde\Lambda,  \nonumber \\ 
\delta_b \, X^{I } = i \Lambda  c    X^{I }-  iX^{I }  \tilde c \ \tilde\Lambda, 
 &&  \delta_b \, X^{I \dagger }
 = i \tilde\Lambda  \tilde c     X^{I \dagger }\  - i  X^{I \dagger }  c \ \Lambda, 
\nonumber \\
\delta_b \, Y^{I } = i\tilde\Lambda   \tilde c    Y^{I }\ -iY^{I }   c \ \Lambda,  &&  
\delta_b \, Y^{I \dagger } = i\Lambda c    Y^{I \dagger } \ -  i Y^{I \dagger }
  \tilde c \ \tilde\Lambda.
  \label{brstnl}
\end{eqnarray}
 Remarkably, the effective action is also found invariant under the
 another set of BRST symmetry (called as anti-BRST transformation) where roles of ghost and anti-ghost fields are interchanged. The anti-BRST transformation is written by
\begin{eqnarray}
\delta_{ab} \,\Gamma_{a} = \nabla_a    \bar c \ \bar\Lambda, && \delta_{ab}\, \tilde\Gamma_{a} =\tilde\nabla_a    
 \tilde{\bar c }\ \tilde{\bar \Lambda}, \nonumber \\
\delta_{ab} \,\bar c  = - \frac{1}{2}{[\bar c , \bar c ]}_ { }\ \bar\Lambda, &&\delta_{ab} \,\tilde{\bar {c}}  = -  \frac{1}{2}  [\tilde{\bar {c}} ,  \tilde {\bar c} ]_{ }\ \tilde{\bar \Lambda}, \nonumber \\
\delta_{ab} \, {c}  = -b \ \bar\Lambda-\frac{1}{2}[\bar c , c ]\ \bar\Lambda, &&\delta_{ab} \,\tilde {  c}  = -\tilde b \ \tilde{\bar \Lambda} -\frac{1}{2}[\tilde {\bar c} , \tilde c ]\ \tilde{\bar \Lambda}, \nonumber \\ 
\delta_{ab} \,b  =-\frac{1}{2}[  \bar c , b ]\ \bar\Lambda +\frac{1}{8}[   [\bar c , \bar c ], c ]\ \bar\Lambda, &&\delta_{ab} \, \tilde b = -\frac{1}{2}
[ \tilde {\bar c} , \tilde b ]\ \tilde{\bar \Lambda}+\frac{1}{8}  [[\tilde {\bar c} , \tilde {\bar c} ], \tilde{  c} ]\ \tilde{\bar \Lambda},  \nonumber \\ 
\delta_{ab} \, X^{I } = i \bar\Lambda \bar c    X^{I }\ -  iX^{I }  \tilde {\bar c} \ \tilde{\bar \Lambda}, 
 && \delta_{ab} \, X^{I \dagger }
 = i \tilde{\bar \Lambda}  \tilde {\bar c}    X^{I \dagger }\  - i  X^{I \dagger } \bar c \ \bar\Lambda, 
\nonumber \\
\delta_{ab} \, Y^{I } = i \tilde{\bar \Lambda} \tilde{\bar c}    Y^{I } \ -iY^{I } \bar  c \ \bar\Lambda,  &&  
\delta_{ab} \, Y^{I \dagger } = i \bar\Lambda  \bar c    Y^{I \dagger }\-  i Y^{I \dagger }
  \tilde{\bar c} \ \tilde{\bar \Lambda}.
\end{eqnarray}
The above BRST and anti-BRST transformations satisfy the following algebra: 
\begin{eqnarray}
\delta_{b}^2=0,\ \ \delta_{ab}^2 =0,\ \ \delta_{b}\delta_{ab} +\delta_{ab} \delta_{b}=0.
\end{eqnarray}
With these BRST and anti-BRST transformations the Lagrangian density (\ref{lag})
can also be expressed  as 
\begin{eqnarray}
{\cal L}_{NL} &=&{\cal L}_c+ \frac{i}{2}\delta_{b}\delta_{ab}\int d^2\theta\   \mbox{Tr} \left[\Gamma_a \Gamma^a -\tilde \Gamma_a\tilde\Gamma^a -i\alpha
\bar c  c  +i\alpha \tilde{\bar c }\tilde c \right]  ,\nonumber\\
&=&{\cal L}_c-\frac{i}{2} \delta_{ab} \delta_{b}\int d^2\theta\  \mbox{Tr}\left[\Gamma_a \Gamma^a -\tilde \Gamma_a\tilde\Gamma^a -i\alpha
\bar c c  +i\alpha \tilde{\bar c }\tilde c \right]  .
\end{eqnarray}

\section{Finite Field Dependent Transformation}
In this section we construct finite field dependent  BRST transformation \cite {jm} of
ABJM theory in ${\cal N}=1$ superspace. To do that we 
first   define two sets of generaric fields as $\Phi^{i }_L (x, \kappa)
 \equiv (\Gamma_{a}, X^I, Y^I, c, \overline{c}, b )$ and $\Phi^{i }_R (x, \kappa)
 \equiv (\tilde\Gamma_{a}, \tilde X^I, \tilde Y^I, \tilde c, \tilde{\overline{c}}, b)$ , here the parameter 
 $\kappa: 0\le \kappa \le 1$. Here   ${\Phi^i}_L   (x, 0 ), {\Phi^i}_R   (x, 0 )$ are the initial fields and
 $ {\Phi^i}_L   (x, 1), {\Phi^i}_R   (x, 1 )$ are the transformed fields. 

The infinitesimal but field dependent BRST transformations can be written as \cite{jm}
\begin{eqnarray}
 \frac{ d}{d \kappa}{\Phi^i}_L   (x, \kappa ) &=& s  {\Phi^i}_L    (x  )\
\epsilon_L [{\Phi}_L   (x,\kappa )], 
\nonumber \\ 
 \frac{ d}{d \kappa}{\Phi^i}_R   (x, \kappa ) &=& s  {\Phi^i}_R    (x  )\
\epsilon_L [{\Phi}_R   (x,\kappa )].
\end{eqnarray}
where $ \epsilon_L  [{\Phi}_L   (x)]$ and $\epsilon_R  [{\Phi}_R   (x)]$
are infinitesimal field dependent parameters.
Now integrating  the above equation from $ \kappa=0$ to $\kappa=1$, we get the FFBRST transformation,
\begin{eqnarray}
 {\Phi^i}_L (x, 1) = {\Phi^i}_L   (x, 0) + s  {\Phi^i}_L    (x) \Theta_L [{\Phi}_L   (x)],
\nonumber \\
{\Phi^i}_R (x, 1) = {\Phi^i}_R   (x, 0) + s  {\Phi^i}_R    (x) \Theta_R [{\Phi}_R   (x)],
\end{eqnarray} 
where finite field dependent parameters are
\begin{eqnarray}
\Theta_L [{\Phi}_L   (x)] &= &\int_0^1 d\kappa\ \epsilon_L [{\Phi}_L   (x,\kappa )],\nonumber\\
\Theta_R [{\Phi}_R   (x)] &= &\int_0^1 d\kappa\ \epsilon_R [{\Phi}_R   (x,\kappa )].\label{para}
\end{eqnarray}
Furthermore these finite parameters are calculated \cite{jm} as,
\begin{eqnarray}
 \Theta_L  [{\Phi}_L   (x)] &=& \epsilon_L  [{\Phi}_L   (x)] \frac{ \exp F_L [{\Phi}_L   (x)]
-1}{F_L [{\Phi}_L   (x)]},
\nonumber \\
 \Theta_R  [{\Phi}_R   (x)] &=& \epsilon_R  [{\Phi}_R   (x)] \frac{ \exp F_R [{\Phi}_R   (x)]
-1}{F_R [{\Phi}_R   (x)]}, 
\end{eqnarray}
where
\begin{eqnarray}
 F_L &=&  \sum_i\frac{ \delta \epsilon_L [\Phi_L (x)]}{\delta
\Phi^i_L (x)} s  \Phi^i_L (x), \nonumber \\
F_R &=&  \sum_i\frac{ \delta \epsilon_R [\Phi_R (x)]}{\delta
\Phi^i_R (x)} s  \Phi^i_R (x). \label{ref}
\end{eqnarray}

Now the FFBRST transformations in the linear gauge are  given by 
\begin{eqnarray}
\delta_b \,\Gamma_{a} = \nabla_a    c \ \Theta_L, && \delta_b\, \tilde\Gamma_{a} =\tilde\nabla_a    
 \tilde c \  \Theta_R, \nonumber \\
\delta_b\,c  = - {[c ,c ]}_ { } \Theta_L, && \delta_b \,\tilde{ {c}}  = -   [\tilde{ {c}} ,  \tilde c ]_{ } \Theta_R, \nonumber \\
\delta_b \,\bar{c}  = b \ \Theta_L, && \delta_b \,\tilde {\bar c}  = \tilde b \  \Theta_R, \nonumber \\ 
\delta_b \,b  =0, &&\delta_b \, \tilde b = 0, \nonumber \\ 
\delta_b \, X^{I } = i  \Theta_Lc    X^{I }  -  iX^{I }  \tilde c  \Theta_R, 
 &&  \delta_b \, X^{I \dagger }
 = i   \Theta_R \tilde c     X^{I \dagger } - i  X^{I \dagger }  c \ \Theta_L, 
\nonumber \\
\delta_b \, Y^{I } = i \Theta_R \tilde c    Y^{I }  -iY^{I }   c \ \Theta_L,  &&  
\delta_b \, Y^{I \dagger } = i \Theta_L c    Y^{I \dagger }\  -  i Y^{I \dagger }
  \tilde c \ \Theta_R.
\end{eqnarray}
and the FFBRST transformations in the non-linear gauge are given by 
\begin{eqnarray}
\delta_b \,\Gamma_{a} = \nabla_a    c \ \Theta_L, && \delta_b\, \tilde\Gamma_{a} =\tilde\nabla_a    
 \tilde c \ \Theta_R, \nonumber \\
\delta_b \,c  = -\frac{1}{2} {[c ,c ]}_ { }\ \Theta_L, && \delta_b \,\tilde{ {c}}  = - \frac{1}{2}  [\tilde{ {c}}  ,  \tilde c ]_{ } \ \Theta_R, \nonumber \\
\delta_b \,\bar{c}  = b \ \Theta_L -\frac{1}{2}[\bar c , c ]\Theta_L, && \delta_b \,\tilde {\bar c}  = \tilde b \ \Theta_R-\frac{1}{2}[\tilde {\bar c} , \tilde c ]\ \Theta_R, \nonumber \\ 
\delta_b \,b  =-\frac{1}{2}[c , b ]\Theta_L - \frac{1}{8}[  [c ,c ] , \bar c ]\Theta_L, &&\delta_b \, \tilde b = -\frac{1}{2}[\tilde c , \tilde b ] \Theta_R - \frac{1}{8}[ [\tilde c , \tilde c ], \tilde{\bar c }]\Theta_R,  \nonumber \\ 
\delta_b \, X^{I } = i \Theta_L c    X^{I }-  iX^{I }  \tilde c \ \Theta_R, 
 &&  \delta_b \, X^{I \dagger }
 = i  \Theta_R \tilde c     X^{I \dagger }\  - i  X^{I \dagger }  c \ \Theta_L, 
\nonumber \\
\delta_b \, Y^{I } = i \Theta_R \tilde c    Y^{I }\  -iY^{I }   c \ \Theta_L,  &&  
\delta_b \, Y^{I \dagger } = i \Theta_L c    Y^{I \dagger } \ -  i Y^{I \dagger }
  \tilde c \ \Theta_R.
\end{eqnarray} 
 The Jacobians for path integral measures  in the expression of generating functionals  are given by 
\begin{eqnarray}
 {\cal D}{\Phi^i}_L    =J_L[{\Phi}_L   (\kappa)] {\cal D}{\Phi^i}_L   (\kappa),
&&
 {\cal D}{\Phi^i}_R    =J_R[{\Phi}_R   (\kappa)] {\cal D}{\Phi^i}_R   (\kappa).
\end{eqnarray}
So, the FFBRST transformations  are  not a symmetry 
of the generating functional.          
Now $J_L[{\Phi}_L   (\kappa)]J_R[{\Phi}_R   (\kappa)]$ can be replaced within the 
functional integral by
$\exp\, ({iS_{1L}[{\Phi}_L   (\kappa)]+iS_{1R}[{\Phi}_R   (\kappa)]})$, if the following 
equations are satisfied, 
\begin{eqnarray}
\int d^2\theta \ \mbox{Tr} \left[\frac{1}{J_L (\kappa )}\frac{d J_L (\kappa )}{d\kappa} -i\frac{dS_{1L} }{d\kappa}\right]   =0, 
\nonumber \\ 
\int d^2\theta \ \mbox{Tr}  \left[\frac{1}{J_R(\kappa )}\frac{d J_R (\kappa )}{d\kappa} -i\frac{dS_{1R}}{d\kappa}\right]   =0. 
\label{mcond}
\end{eqnarray}
The infinitesimal changes in Jacobian's are given by 
 \begin{eqnarray} 
 \frac{1}{J_L (\kappa )}\frac{d J_L (\kappa )}{d\kappa} &=& 
 -\int d^2\theta \ \mbox{Tr}  \  \mathcal{A}_L , 
\nonumber \\ 
  \frac{1}{J_R(\kappa)}\frac{dJ_R(\kappa)}{d\kappa}&=& 
 -\int d^2\theta \ \mbox{Tr}  \  \mathcal{A}_R, \label{jaceva}
\end{eqnarray}
where explicit expressions for $\mathcal{A}_L$ and $\mathcal{A}_R$ are given by
\begin{eqnarray}
 \mathcal{A}_L &=& \left[  s  \Gamma_a (x, \kappa) \frac{ \delta \epsilon_L [\Phi_L (x, k)]}{\delta
\Gamma_a (x, k)}   -s  c (x, k)
\frac{ \delta \epsilon_L [\Phi_L (x)]}{\delta
c  (x, k)} \right. \nonumber \\ &&\left. - s  \overline c  (x, k)
\frac{ \delta \epsilon_L [\Phi_L (x, k)]}{\delta
\overline c  (x, k)}    +s b (x, k)
\frac{ \delta \epsilon_L [\Phi_L (x, k)]}{\delta
b (x, k)}
\right. \nonumber \\ &&\left. - s X^I  (x, k)
\frac{ \delta \epsilon_L [\Phi_L (x, k)]}{\delta
X^I  (x, k)}    +s Y^I (x, k)
\frac{ \delta \epsilon_L [\Phi_L (x, k)]}{\delta
Y^I (x, k)}
\right]  , \nonumber \\
  \mathcal{A}_R &=&  \left[  s  \tilde{\Gamma}_a (x,\kappa) \frac{ \delta \epsilon_R [\Phi_R (x, k)]}{\delta
 \tilde{\Gamma}_a  (x, k)}   -s  \tilde c  (x, k)
\frac{ \delta \epsilon_R [\Phi_R (x)]}{\delta
\tilde c (x, k)}  \right. \nonumber \\ &&\left.  - s  \tilde{\overline c}  (x, k)
\frac{ \delta \epsilon_R [\Phi_R (x, k)]}{\delta
\tilde{\overline c} (x, k)}   +s \tilde b(x, k)
\frac{ \delta \epsilon_R [\Phi_R (x, k)]}{\delta
\tilde b (x, k)}\right. \nonumber \\ &&\left. - s X^{I\dag}  (x, k)
\frac{ \delta \epsilon_R [\Phi_R (x, k)]}{\delta
X^{I\dag} (x, k)}    +s Y^{I\dag} (x, k)
\frac{ \delta \epsilon_R [\Phi_R(x, k)]}{\delta
Y^{I\dag} (x, k)}
\right].
\end{eqnarray}
Here we note that the conditions (\ref{mcond}) provide us liberty to replace the
Jacobians of path integral measure by the exponential of local functional within functional measure.
Hence, the Jacobians amount a precise change in effective action of  generating functional . One can also arrive at the 
same conclusion following the work in Ref.  \cite{lav,lav1}.

\section{ Relating Different Gauges} 
We will now use FFBRST to relate  the generating functional in the linear gauge to the
generating functional in the non-linear gauge. 
If the gauge fixing condition in the 
linear gauge is denoted by    $G_{1L}[\Gamma_a], G_{1R}[ \tilde \Gamma_a]$ 
and the gauge fixing condition in the non-linear gauge is denoted by 
 $G_{2L}[\Gamma_a],  G_{1R}[ \tilde\Gamma_a]$, 
then, the linear BRST transformations of $G_{1L}[\Gamma_a], G_{1R}[ \tilde\Gamma_a]$ are denoted by $sG_{1L}, sG_{1R}$ and the 
the non-linear BRST transformations of  $G_{2L}[\Gamma_a],  G_{1R}[ \tilde\Gamma_a]$ are denoted by $sG_{2L}, s G_{2R}$. 
We define the infinitesimal field dependent parameter as follows  
\begin{eqnarray}
\epsilon_L [\Phi_L] &=&  
i\gamma\int d^2\theta \ \mbox{Tr} \  \left[ \overline c  \left( G_{ 1L} - G_{ 2L}\right)\right]  ,
\nonumber \\
\epsilon_R [\Phi_R] &=&    -i\gamma\int d^2\theta \ \mbox{Tr}\  \left[\tilde{\overline c}  \left( G_{ 1R} - G_{ 2R}\right)\right]  , 
\end{eqnarray}
where $\gamma$ is an arbitrary constant parameter.

Using definition given in (\ref{jaceva}), the change in Jacobian's can be calculated as follows, 
\begin{eqnarray}
  \frac{1}{J_L}\frac{dJ_L}{d\kappa} &=&i\gamma\int d^2\theta \ \mbox{Tr} \  \left[ b G_{ 1L} - b G_{ 2L}
 - ( sG_{ 1L} - sG_{ 2L})\bar c \right]  ,\nonumber\\
&=&i\gamma\int d^2\theta \ \mbox{Tr} \  \left[ b G_{ 1L} - b G_{2L}
+\bar c ( sG_{ 1L} - sG_{2L})\right]  ,
 \nonumber \\
  \frac{1}{J_R}\frac{dJ_R}{d\kappa} &=&- i\gamma\int d^2\theta \ \mbox{Tr} \ \left[ \tilde b G_{ 1R} -\tilde b G_{ 2R}
-( sG_{ 1R} - sG_{ 2R})\tilde{\bar c} \right]  ,\nonumber\\
&=&- i\gamma\int d^2\theta \ \mbox{Tr} \   \left[ \tilde b G_{ 1R} - \tilde b G_{2R}
+\tilde {\bar c} ( sG_{ 1R} - sG_{2R})\right]  .
\end{eqnarray}
Furthermore, the local functionals $S_{1L}$ and $S_{1R}$ involved in  the Jacobians 
are defined as
\begin{eqnarray}
S_{1L} &=&\int d^2\theta \ \mbox{Tr} \   [\xi_{1L} (\kappa) b G_{ 1L}+ 
\xi_{2L} (\kappa) b G_{ 2L} \nonumber \\ && +\xi_{3L}(\kappa) \overline c 
s G_{1L} +\xi_{4L}(\kappa)\overline c  sG_{ 2L} ]  ,
\nonumber \\
S_{1R} &=& \int d^2\theta \ \mbox{Tr} \   [\xi_{1R} (\kappa) \tilde b G_{ 1R}+ 
\xi_{2R} (\kappa) \tilde b G_{ 2R} \nonumber \\ && +\xi_{3R}(\kappa) \tilde{\overline c}
s G_{1R} +\xi_{4R}(\kappa)\tilde{\overline c } sG_{ 2R} ]  ,
\end{eqnarray}
where $\xi_{iL}, \xi_{iR}, (i=1,2,3,4)$ are $\kappa$ dependent 
arbitrary parameters which satisfy the following initial boundary conditions, 
$
\xi_{iL} (\kappa =0) = \xi_{iR}(\kappa =0)=0
$. 
As all the fields depend on $\kappa$, so we can write  
\begin{eqnarray}
\frac{dS_{1L}}{d\kappa}&=&\int d^2\theta \ \mbox{Tr} \   [\xi'_{1L}  b G_{ 1L} 
+\xi_{1L} b 
s G_{ 1L}\epsilon_L   \nonumber \\ 
 &&+\xi_{2R} b 
s G_{2L} \epsilon_L +\xi'_{3L} 
\overline c 
s G_{1L} -\xi_{3L} b 
s G_{1L}\epsilon_L  \nonumber\\
 &&+\xi'_{4L} \overline c  sG_{2_L}
  -\xi_{4L} b 
s G_{2L} \epsilon_L +  \xi'_{2L}  b G_{2L}]  ,\nonumber\\
 &=&\int d^2\theta \ \mbox{Tr} \   [\xi'_{1L}  b G_{ 1L} +  \xi'_{2L}  b G_{ 2L}\nonumber \\ & 
 &+\xi'_{3L} 
\overline c 
s G_{ 1L} +\xi'_{4L} \overline c  sG_{ 2L} \nonumber\\
&&+
(\xi_{1L} -\xi_{3L} )b 
s G_{ 1L}\epsilon_L +(\xi_{2L} -\xi_{4L}) b 
s G_{2L} \epsilon_L]  , 
\nonumber \\ 
\frac{dS_{1R}}{d\kappa}&=&\int d^2\theta \ \mbox{Tr} \  [\xi'_{1R}  \tilde b G_{ 1R} 
+\xi_{1R} \tilde b 
s G_{ 1R}\epsilon_R   \nonumber \\ 
 &&+\xi_{2R} \tilde b 
s G_{2R} \epsilon_R +\xi'_{3R} 
\tilde{\overline c }
s G_{1R} -\xi_{3R} \tilde b 
s G_{1R}\epsilon_R  \nonumber\\
 &&+\xi'_{4R} \tilde{\overline c}  sG_{2_R}
  -\xi_{4R} \tilde b 
s G_{2R} \epsilon_R +  \xi'_{2R}  \tilde b G_{2R}]  ,\nonumber\\
 &=&\int d^2\theta \ \mbox{Tr} \ [\xi'_{1R} \tilde b G_{ 1R} +  \xi'_{2R}  \tilde b G_{ 2R}\nonumber \\ & 
 &+\xi'_{3R} 
\tilde{\overline c} 
s G_{ 1R} +\xi'_{4R} \tilde{\overline c} sG_{ 2R} \nonumber\\
&&+
(\xi_{1R} -\xi_{3R} )\tilde b
s G_{ 1R}\epsilon_R +(\xi_{2R} -\xi_{4R})\tilde b 
s G_{2R} \epsilon_R]  .
\end{eqnarray}

The Jacobians of path integral measure can be written as  $\exp\, ({iS_{1L}+ iS_{1R}})$, when the following equations are satisfied, 
\begin{eqnarray}
&&\int d^2\theta \ \mbox{Tr} \  \left[ 
(\xi'_{1L}-\gamma)  b G_{ 1L}+( \xi'_{2L}  +\gamma ) b G_{ 2L}
\right.\nonumber\\
&&\left. + ( \xi'_{3L} -\gamma )
\overline c  sG_{1L}  + (\xi'_{4L} +\gamma ) \overline c  sG_{ 2L} 
\right.\nonumber\\
&&\left.+(\xi_{1L} -\xi_{3L} )b  sG_{1L}\epsilon_L  +(\xi_{2L} -\xi_{4L}) b sG_{2L}\epsilon_L  \right]  =0, 
\nonumber \\
&&\int d^2\theta \ \mbox{Tr} \   \left[ 
(\xi'_{1R}+\gamma) \tilde{b}  G_{ 1R}+( \xi'_{2R}  -\gamma ) \tilde bG_{ 2R}
\right.\nonumber\\
&&\left. + ( \xi'_{3R} +\gamma )
\tilde{\overline c}  sG_{1R}  + (\xi'_{4R} -\gamma ) \tilde{\overline c}  sG_{ 2R} 
\right.\nonumber\\
&&\left.+(\xi_{1R} -\xi_{3R} )\tilde b sG_{1R}\epsilon_R  +(\xi_{2R} -\xi_{4R}) \tilde b
 sG_{2R}\epsilon_R  \right]  =0.
\end{eqnarray}
 Equating the coeffiecients of the above expressions, and setting  $\gamma =1$,
we get
$
\xi_{1L} = -\xi_{1R} = \kappa, \, \xi_{2L} = -\xi_{2R} = -\kappa, \, \xi_{3L} = -\xi _{3R} = \kappa, \, 
\xi_{4L} = -\xi_{4R} = -\kappa
$.
Now, if we add  $ S_1 = S_{1L} (\kappa =1 ) + S_{1R} (\kappa =1 )$ to the original action in the non-linear 
gauge, we obtain the action in the linear gauge within a functional integral. 
So,   under the FFBRST transformations the  generating functional 
in the non-linear gauge  transforms to the generating functional 
 in the linear gauge.  
Similar computations can also been made  following the work in Ref. \cite{lav,lav1} to show that the
FFBRST transformation amounts finite change in gauge-fixing fermion
of the path integral.

\section{Conclusion}
In this paper we analysed the FFBRST transformations for the ABJM theory in ${\cal N}=1$ superspace. 
We first have discussed the BRST for the ABJM theory in ${\cal N}=1$ superspace. Then we have integrated the infinitesimal 
parameter in the BRST transformations to obtain the FFBRST transformations. As the ABJM theory contains two
Chern-Simons terms, we have constructed two finite parameters in the FFBRST transformations. These parameters are only mixed 
due to the matter terms. The BRST transformations of this theory have been studied 
in both linear as well as non-linear gauges.  
After analysing   both the linear and non-linear BRST transformations, a finite field
dependent version of these 
transformations has been developed. It has been  shown that these two gauges can be related to each other via  
FFBRST transformations.

Multiple D2-brane action has been derived from a multiple M2-brane action by 
means of a novel Higgs mechanism \cite{d2,sxsw, xswd, d21}. In this mechanism 
a vacuum expectation value is given to a scalar field which  breaks  the gauge group   $U(N) \times U(N)$
 down to its diagonal subgroup. The theory thus obtained is  the Yang-Mills theory coupled to matter fields. 
It would be interesting to start with a gauge fixed ABJM theory in ${\cal N}=1$ superspace and 
use the novel Higgs mechanics to obtain the Yang-Mills theory coupled to matter fields. 
It would also be interesting to study the FFBRST transformations of the ABJM theory and the 
FFBRST transformations of the theory obtained after using the novel Higgs mechanics.  
It is  expected that the FFBRST transformations for the ABJM theory will reduce to the 
 the FFBRST transformations for the Yang-Mills theory coupled to matter fields. 
 
There is a dual symmetry to the BRST symmetry called the anti-BRST symmetry \cite{ an1, an2}. 
The finite field version of anti-BRST (anti-FFBRST) symmetry has also been studied \cite{antifbrst,alex}. 
It would be interesting to study this symmetry for the ABJM theory in ${\cal N}=1$ superspace. 
Furthermore, the ABJM theory in presence of a boundary has also been analysed \cite{abb}. 
In this theory new boundary degrees of freedom have to be added to make this theory 
gauge invariant. It would be interesting to analyse the FFBRST and anti-FFBRST symmetry 
for this ABJM theory in presence of a boundary. 
These transformations can be used to relate the generating functionals 
in case of ABJM theory in presence of a boundary.


\begin{thebibliography}{99}

 \bibitem{1}A. Gustavsson, JHEP. 0804, 083 (2008)
 \bibitem{2}J. Bagger and N. Lambert, JHEP. 0802, 105 (2008)
\bibitem{3}J. Bagger and N. Lambert, Phys. Rev. D77, 065008 (2008)
 \bibitem{4}M. A. Bandres, A. E. Lipstein and J. H. Schwarz, JHEP. 0809, 027 (2008)
\bibitem{5}E. Antonyan and  A. A. Tseytlin, Phys. Rev. D79, 046002 (2009)
\bibitem{abjm}O. Aharony, O. Bergman, D. L. Jafferis and J. Maldacena,
 JHEP. 0810, 091 (2008)
 \bibitem{ab}
 M. A. Bandres, A. E. Lipstein and J. H. Schwarz, JHEP. 0809, 027 (2008)
 \bibitem{ab1}
 M. Schnabl and Y.  Tachikawa, JHEP. 1009, 103 (2010)
 \bibitem{ab2}
 E. Antonyan and A. A. Tseytlin, Phys. Rev. D79,  046002 (2009)
 \bibitem{abjm2}O-Kab Kwon, P. Oh and  J. Sohn, JHEP.  0908, 093 (2009)
 \bibitem{6} M. Benna, I. Klebanov, T. Klose and  M. Smedback, JHEP. 0809, 072  (2008)
 \bibitem{7} A. Mauri and  A. C.  Petkou, Phys. Lett. B666
527 (2008) 
 \bibitem{8} S. Cherkis, and C. Samann   Phys. Rev. D78,   066019 (2008)
 \bibitem{ahs}
I. L. Buchbinder, E. A. Ivanov, O. Lechtenfeld, N. G. Pletnev, I. B. Samsonov and B.M. Zupnik, JHEP. 0903, 096 (2009) 
 
\bibitem{mir4}M. Faizal, Mod. Phys. Lett. A27, 1250147 (2012) 

\bibitem{abm}
M. Faizal,  Phys. Rev. D84, 106011 (2011)

 
\bibitem{jm} S. D. Joglekar and B. P. Mandal,  Phys. Rev. D 51,  1919 (1995)
 \bibitem{antifbrst} S. Upadhyay, S. K. Rai and B. P. Mandal,  J. Math. Phys. 52, 022301 (2011)
 \bibitem{lav} P. Lavrov, O. Lechtenfeld, Phys. Lett. B 725, 382 (2013)
\bibitem{lav1} I. A. Batalin, P. M. Lavrov and I. V. Tyutin, arXiv: 1405.262 [hep-th]; arXiv: 1404.4154 [hep-th]; 
arXiv: 1406.4695 [hep-th]; arXiv: 1405.7218 [hep-th]
\bibitem{ffb} S. D. Joglekar and A. Misra,  Int. J. Mod. Phys. A 15, 1453 (2000) 
\bibitem{ffb1} S. D. Joglekar and B. P. Mandal,  Int. J. Mod. Phys. A 17, 1279 (2002)

\bibitem{ffb2} R. Banerjee and B. P. Mandal,  Phys. Lett. B 488, 27 (2000)
\bibitem{ffb3} S. K. Rai and B. P. Mandal,  Eur. Phys. J. C  63, 323 (2009) 
\bibitem{sudm} S. Upadhyay  and B. P. Mandal, Annals. Phys. 327, 2885  (2012)
\bibitem{sudm1} S. Upadhyay  and B. P. Mandal, Eur. Phys. J. C 72, 2065 (2012)
 \bibitem{sudm3}  S. Upadhyay    and B. P. Mandal, Mod. Phys. Lett. A 25, 3347 (2010)
 
 \bibitem{ff1} S. K. Rai and  B. P. Mandal, Int. J. Theor. Phys. 52, 3512 (2013)

 \bibitem{sud} S. Upadhyay, Phys. Lett. B 727, 293 (2013)
 \bibitem{sud1} S. Upadhyay, Annals.  Phys. 340, 110  (2014)
  \bibitem{sud2} S. Upadhyay, Annals.  Phys.  344, 290 (2014)
  \bibitem{sud3} S. Upadhyay and B. P. Mandal,  Prog. Theor. Exp. Phys. 053B04  (2014)
  
\bibitem{rbs1} R. Banerjee  and S. Upadhyay, Phys. Lett. B 734, 369 (2014) 
 \bibitem{rbs} R. Banerjee, B. Paul and S. Upadhyay,  Phys. Rev. D 88, 065019 (2013)
  \bibitem{gz} S. Upadhyay  and B. P. Mandal, Euro. Phys. Lett. 93, 31001 (2011); AIP Conf. Proc. 1444, 213 (2012) 
  \bibitem{lav2} P. Lavrov, O. Lechtenfeld and A. Reshetnyak, JHEP 1110, 043 (2011)
  \bibitem{lav3}P. Lavrov and O. Lechtenfeld, Phys. Lett. B 725, 386 (2013)
 \bibitem{m5}S. P. Chowdhury and S. Chakrabortty and K. Ray, Phys. Lett. B703, 172 (2011) 
 \bibitem{m5fgc} M. A. Ganjali, JHEP. 0903, 064 (2009)
 \bibitem{gygfvjk} P. M.  Ho, Y.  Imamura, Y. Matsuo and S. Shiba, JHEP. 0808, 014 (2008) 
 \bibitem{m512} A.  Gustavsson, JHEP. 0911, 071 (2009) 
\bibitem{fs} M. Faizal, B. P. Mandal and S. Upadhyay, Phys. Lett. B 721, 159 (2013)
 

\bibitem{d2} S. Mukhi and C. Papageorgakis, JHEP. 0805, 085 (2008) 
\bibitem{sxsw} T. Li, Y. Liu and D. Xie, Int. J. Mod. Phys. A24, 3039 (2009) 
\bibitem{xswd}   Y. Pang and T. Wang, Phys. Rev. D78, 125007 (2008) 
\bibitem{d21} P. M. Ho, Y. Imamura and Y. Matsuo,  JHEP. 0807, 003 (2008) 
\bibitem{an1} G. Curci and R. Ferrari, Phys. Lett. B 63, 91 (1976)
\bibitem{an2} I. Ojima, Prog. Theor. Phys. 64, 625 (1979)
\bibitem{alex} P. Y. Moshin and A. Reshetnyak, arXiv:1405.0790 [hep-th]; arXiv:1405.7549[hep-th];
arXiv:1406.5086 [hep-th] 
\bibitem{abb}M. Faizal and D. J. Smith, Phys. Rev. D85, 105007 (2012)
\end{thebibliography}
\end{document}